\documentclass{easychair}

\usepackage{latexsym}
\usepackage{amsmath}
\usepackage{stmaryrd}
\usepackage{graphicx}
\usepackage{pict2e}
\usepackage{tikz}
\usepackage{pgfplots}
\usepackage{cite}
\usepackage{booktabs}
\usepackage{adjustbox}
\usepackage{marvosym}
\newcommand{\cref}[2]{\href{#1}{\color{blue}#2}}
\newcommand{\hcref}[1]{\cref{#1}{\small\tt #1}}
\usepackage{inconsolata}
\usepackage[T1]{fontenc}
\usepackage[textwidth=4cm]{todonotes}
\usepackage{longtable}
\usepackage{tabularx}

\newcommand{\kissat}{{\sffamily\scshape kissat}}
\newcommand{\Kissat}{{\sffamily\scshape  Kissat}}
\newcommand{\cms}{{\sffamily\scshape  CryptoMiniSat}}
\newcommand{\cmsshort}{{\sffamily\scshape  CMS}}

\newcommand{\Cms}{{\sffamily\scshape  CryptoMiniSat}}

\newcommand{\buddy}{{\sffamily\scshape  buddy}}

\newcommand{\tbuddy}{{\sffamily\scshape  tbuddy}}
\newcommand{\Tbuddy}{{\sffamily\scshape  Tbuddy}}
\newcommand{\tbsat}{{\sffamily\scshape  tbsat}}
\newcommand{\minisat}{{\sffamily\scshape  MiniSat}}
\newcommand{\Tbsat}{{\sffamily\scshape  Tbsat}}
\newcommand{\Veripb}{{\sffamily\scshape  VeriPB}}
\newcommand{\drattrim}{{\sffamily\scshape  drat-trim}}
\newcommand{\Drattrim}{{\sffamily\scshape  Drat-trim}}
\newcommand{\fratrs}{{\sffamily\scshape  frat-rs}}

\newcommand{\lrat}{{\sffamily\scshape  lrat}}
\newcommand{\lratcheck}{{\sffamily\scshape  lrat-check}}

\newcommand{\makenode}[1]{{\mathbf #1}}
\newcommand{\nodeu}{\makenode{u}}
\newcommand{\nodev}{\makenode{v}}
\newcommand{\nodew}{\makenode{w}}

\newcommand{\leafzero}{\makenode{T}_0}

\newcommand{\boolxor}{\oplus}

\newcommand{\tautology}{1}
\newcommand{\nil}{0}

\newcommand{\trust}[1]{\dot {#1}}
\newcommand{\tleafzero}{\trust{\makenode{T}}_0}
\newcommand{\lit}{\ell}

\newcommand{\fname}[1]{\mbox{\small\sf #1}}

\newcommand{\var}{\fname{Var}}
\newcommand{\phase}{\fname{Phase}}

\newcommand{\node}[1]{\mathbf{#1}}

\newcommand{\psum}{{\Sigma}}

\title{Proof Generation for CDCL Solvers Using \\ Gauss-Jordan Elimination}

\author{
  Mate Soos\inst{1}
  \and
  Randal E. Bryant\inst{2}~\thanks{Supported by the U. S. National Science Foundation under grant CCF-2108521}
}

\institute{
  National University of Singapore\\
    {\tt soos.mate@gmail.com}
  \and
    Computer Science Department \\
    Carnegie Mellon University, Pittsburgh, PA, United States\\
    \email{Randy.Bryant@cs.cmu.edu}
}

\authorrunning{M. Soos and R. E. Bryant}

\titlerunning{Proof Generation for CDCL Solvers Using Gauss-Jordan Elimination}

\begin{document}

\maketitle

\begin{abstract}
Traditional Boolean satisfiability (SAT) solvers based on the
conflict-driven clause-learning (CDCL) framework fare poorly on
formulas involving large numbers of parity constraints.  The \cms{}
solver augments CDCL with Gauss-Jordan elimination to greatly improve
performance on these formulas.  Integrating the \tbuddy{}
proof-generating BDD library into \cms{} enables it to generate
unsatisfiability proofs when using Gauss-Jordan elimination.  These
proofs are compatible with standard, clausal proof frameworks.

\end{abstract}

\section{Introduction}

Consider Boolean formulas over a set of variables $X = \{x_1, x_2,
\ldots, x_n\}$.  A $k$-way {\em parity constraint}, also known as an {\em XOR
constraint}, is an equation of the form:
\begin{eqnarray}
x_{i_1} \boolxor x_{i_2} \boolxor \cdots \boolxor x_{i_k} & = & p \label{eqn:parity}
\end{eqnarray}
where the phase $p$ can be 1 (odd parity) or 0 (even parity.)

Although Boolean satisfiability (SAT) solvers based on conflict-driven
clause learning (CDCL) have made steady improvements over the years,
they fare poorly when the formula contains large numbers of parity
constraints.

As an example, Urquhart devised a family of unsatisfiable formulas, consisting entirely of parity constraints,
where both the number of variables and the number of clauses scale
quadratically with the size parameter $m$, but any resolution proof of
unsatisfiability must scale exponentially in $m$~\cite{Urquhart:1987}.
The smallest instance of this benchmark, having $m=3$, consists of 153 variables
and 408 clauses, encoding 102 three-way parity constraints.
\Kissat{}, a state-of-the-art CDCL solver~\cite{biere-kissat-2020}, fails to terminate after
running on this formula for 16 hours, even with proof generation
disabled.  On the other hand, by viewing constraints of the form of Equation~\ref{eqn:parity} as
linear equations over integers modulo 2, applying
Gaussian elimination to an unsatisfiable set of parity constraints yields the
infeasible equation $0 = 1$ in polynomial time.

Several CDCL solvers have been augmented with constraint solvers that
can apply Gauss-Jordan elimination to parity
constraints~\cite{gocht:aaai:2021,han:cav:2012,laitinen:TAI:2012,crypto}.
These solvers combine traditional clausal reasoning with parity
reasoning to improve their performance on both satisfiable and
unsatisfiable formulas.  They operate by
first detecting the parity constraints encoded in the input formula
and delegating the parity constraints to the parity
reasoning component, while retaining the remaining clauses for the
clausal reasoning component.  During execution, the two components reason about their respective portions of the formula
and
coordinate via unit propagation and conflict detection.  This mechanism is sometimes referred to as ``CDCL(T),''
reflecting its similarity to
the handling of multiple theories
by SMT solvers~\cite{demoura:tacas:2008}.
This approach can be extremely effective. For example, version
5.8.0 of \cms{} running on the Urquhart formula with $m=3$ detects that the
formula is unsatisfiable in just one second.  Even scaling to $m=40$,
a formula that is over 200 times larger, with 33,120 variables and
88,320 clauses, the program can detect that the formula is unsatisfiable in less than five minutes.

\subsection{Proofs of Unsatisfiability for SAT Solvers}
Recent generations of CDCL SAT solvers can produce a proof of
unsatisfiability when they encounter an unsatisfiable formula.  Such a
proof provides an independently checkable confirmation that the
formula is truly unsatisfiable.  Since 2016, entrants in the main track of the annual
SAT competition receive credit for an unsatisfiable result only when
1) they produce a DRAT proof~\cite{wetzler14_drattrim}, and 2) this
proof is successfully checked by a standard proof checker.
Unsatisfiability proofs also ensure the
integrity of the formal verification tools and mathematical proofs
that employ SAT solving.  This is especially important when the results
are intended to provide assurances in high risk environments, such as
aerospace, transportation, and cybersecurity.

Until now, generating clausal proofs while employing parity reasoning
has been a major challenge.  Standard clausal proof frameworks, such as DRAT, are well
matched to CDCL solvers, but there has been limited success
generating DRAT proofs while employing Gaussian or Gauss-Jordan elimination.
As a result, parity reasoning has not been used by entrants in the
main track of the SAT competitions in recent years.\footnote{The ``nolimits'' track
allows for solvers to state that a formula is unsatisfiable without
proof, and hence solvers with Gauss-Jordan elimination have continued
to participate in this track.}

Given their inability to generate clausal proofs when using
Gauss-Jordan elimination, most current SAT solvers disable parity
reasoning when they are directed to produce proofs and instead rely
purely on CDCL\@.  In this mode, they fare no better than \kissat{} on
formulas containing parity constraints, including the Urquhart
formulas.

\subsection{Related Work}

Some have proposed the adoption of other proof frameworks to enable
proof generation by parity constraint solvers.  For example, the
\Veripb{} proof checker verifies proofs expressed in a logic of
pseudo-Boolean constraints~\cite{gocht:veripb:2020}.  Using this
framework as a target, Gocht and N\"ordstrom were able to integrate a
Gauss-Jordan elimination solver into the \minisat{} solver and have it
generate proofs of
unsatisfiability~\cite{gocht:aaai:2021,DBLP:conf/cav/SoosGM20}.  In a
similar vein, Barnett and Biere proposed a proof framework based on
binary decision diagrams (BDDs), where each proof step is guaranteed 
to be checkable with polynomial complexity~\cite{barnett:cade:2021}.
Their framework could also enable proof generation from parity
constraints.  Getting the SAT community to adopt a new proof framework would
require establishing and documenting a new set of file formats and creating a
collection of proof checkers, preferably including ones that have been
formally verified.  Our approach has the advantage that it builds on
the well-established DRAT framework.

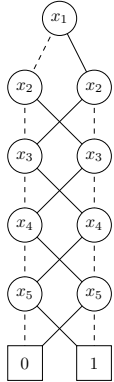
\begin{figure}
\centering{
\scalebox{0.65}{%
\begin{tikzpicture}
\definecolor{fillcolor}{RGB}{255,255,255}
\definecolor{highcolor}{RGB}{0,0,0}
\definecolor{lowcolor}{RGB}{0,0,0}
\definecolor{neutralcolor}{RGB}{0,0,0}
\definecolor{pathcolor}{RGB}{0,0,0}
\definecolor{background}{RGB}{225,225,225}
\draw (1.06,7.41) [thin,highcolor] -- (1.77,6.00);
\draw [thin,lowcolor,dashed] (1.06,7.41) -- (0.35,6.00);
\draw (1.77,6.00) [thin,highcolor] -- (0.35,4.59);
\draw [thin,lowcolor,dashed] (1.77,6.00) -- (1.77,4.59);
\draw (0.35,6.00) [thin,highcolor] -- (1.77,4.59);
\draw [thin,lowcolor,dashed] (0.35,6.00) -- (0.35,4.59);
\draw (1.77,4.59) [thin,highcolor] -- (0.35,3.18);
\draw [thin,lowcolor,dashed] (1.77,4.59) -- (1.77,3.18);
\draw (0.35,4.59) [thin,highcolor] -- (1.77,3.18);
\draw [thin,lowcolor,dashed] (0.35,4.59) -- (0.35,3.18);
\draw (1.77,3.18) [thin,highcolor] -- (0.35,1.77);
\draw [thin,lowcolor,dashed] (1.77,3.18) -- (1.77,1.77);
\draw (0.35,3.18) [thin,highcolor] -- (1.77,1.77);
\draw [thin,lowcolor,dashed] (0.35,3.18) -- (0.35,1.77);
\draw (1.77,1.77) [thin,highcolor] -- (0.35,0.35);
\draw [thin,lowcolor,dashed] (1.77,1.77) -- (1.77,0.35);
\draw (0.35,1.77) [thin,highcolor] -- (1.77,0.35);
\draw [thin,lowcolor,dashed] (0.35,1.77) -- (0.35,0.35);
\draw [thin,fill=fillcolor,draw=neutralcolor] (1.06,7.41) circle [radius=0.35];
\node at (1.06,7.41) {$x_1$};
\draw [thin,fill=fillcolor,draw=neutralcolor] (0.35,6.00) circle [radius=0.35];
\node at (0.35,6.00) {$x_2$};
\draw [thin,fill=fillcolor,draw=neutralcolor] (1.77,6.00) circle [radius=0.35];
\node at (1.77,6.00) {$x_2$};
\draw [thin,fill=fillcolor,draw=neutralcolor] (0.35,4.59) circle [radius=0.35];
\node at (0.35,4.59) {$x_3$};
\draw [thin,fill=fillcolor,draw=neutralcolor] (1.77,4.59) circle [radius=0.35];
\node at (1.77,4.59) {$x_3$};
\draw [thin,fill=fillcolor,draw=neutralcolor] (0.35,3.18) circle [radius=0.35];
\node at (0.35,3.18) {$x_4$};
\draw [thin,fill=fillcolor,draw=neutralcolor] (1.77,3.18) circle [radius=0.35];
\node at (1.77,3.18) {$x_4$};
\draw [thin,fill=fillcolor,draw=neutralcolor] (0.35,1.77) circle [radius=0.35];
\node at (0.35,1.77) {$x_5$};
\draw [thin,fill=fillcolor,draw=neutralcolor] (1.77,1.77) circle [radius=0.35];
\node at (1.77,1.77) {$x_5$};
\draw [thin,fill=fillcolor,draw=neutralcolor] (0.00,0.00) rectangle (0.71,0.71);
\node at (0.35,0.35) {$0$};
\draw [thin,fill=fillcolor,draw=neutralcolor] (1.41,0.00) rectangle (2.12,0.71);
\node at (1.77,0.35) {$1$};
\end{tikzpicture}
}
}
\caption{BDD Representation of a 5-variable parity constraint.  The representation is linear in the number of variables for any ordering of the BDD variables.}
\label{fig:parity:bdd}
\end{figure}

In prior work, we devised a fundamentally different approach for
generating unsatisfiability proofs when reasoning about {\em
  pseudo-Boolean} formulas, of which parity constraints are a special
case~\cite{bryant:tacas:2022}.  We do so by combining Gaussian
elimination with a proof generator based on BDDs.  BDDs are especially
effective for reasoning about parity constraints---the BDD
representation of a $k$-way constraint has only $2k-1$ nonterminal
nodes regardless of the BDD variable ordering.  As an example, Figure~\ref{fig:parity:bdd} shows the BDD representation of the parity constraint
\begin{eqnarray*}
x_1 \boolxor x_2 \boolxor x_3 \boolxor x_4 \boolxor  x_5 & = & 1
\end{eqnarray*}
The proof generator
uses the ability to introduce extension variables into DRAT proofs, in
the style of extended resolution~\cite{kullman:dam:1999,Tseitin:1983}.
An extension variable is added for each BDD node generated, describing
the logical relation between the node and its associated variable and
children nodes~\cite{ebddres,Jussila:2006,bryant:tacas:2021}.  Each
recursive step of a BDD operation is justified by a short sequence of
DRAT proof steps, with the effect that, when the Gaussian elimination
solver detects an infeasible parity constraint, the overall proof
consists of a sequence of implication-preserving clauses terminating
with the empty clause.  On the other hand, our BDD-based solver does
not do well for many of the formulas that are readily handled by CDCL
solvers, and a third class of problems requires both clausal and
parity reasoning to detect that a formula is unsatisfiable.
Integrating the proof generating capabilities of BDDs into a CDCL(T) solver
expands the range of
formulas for which it can generate unsatisfiability proofs.

Philip and Rebola-Pardo devised a method to generate DRAT proofs when
manipulating parity constraints~\cite{philipp:lai:2016}.  Their
approach maintains clausal representations of parity constraints,
using auxiliary variables to ensure that the representation of a
$k$-ary constraint uses $O(k)$ clauses.  With each step of a
constraint solver, their method generates a proof that the clausal
encoding of the newly generated constraint is logically implied by the
encodings of the argument constraints.  The auxiliary variables in the
new constraint become extension variables in the proof.
Operationally, this approach is similar to proof-generating BDD-based
solvers, where extension variables are added for every BDD node
generated.  Their work is specialized to parity constraints, whereas
ours can be applied to other classes of pseudo-Boolean
constraints~\cite{bryant:tacas:2022}.  Gocht and N\"ordstrom
incorporated this method into their extended version of
\minisat{}~\cite{gocht:aaai:2021}.  Although they were able to
generate proofs for benchmark circuits that would be intractable for
pure CDCL solvers, they found that the checking time did not scale
very well.  We speculate that this is due to the poor performance of
the standard checker \drattrim{}~\cite{wetzler14_drattrim} on formulas with large
numbers of extension variables, a shortcoming that we also experienced.

\subsection{Contributions}
In this work, we describe our recent experience integrating the
proof-generating capabilities of BDDs into the Gauss-Jordan (G-J)
solver component of \cms{}, a highly tuned CDCL(T) solver.  We do so using \tbuddy{}~\cite{bryant:fmcad:2022}, a
proof-generating BDD library that was implemented by extending the
\buddy{}~\cite{buddy} BDD package.\footnote{\Tbuddy{} is available at \hcref{https://github.com/rebryant/tbuddy-artifact}.}
\Tbuddy{} generates proof steps to justify the unit propagations and
conflicts inferred during G-J elimination.  This capability was added
via a loose integration, with only one new data structure added to \cms{}.

A bigger challenge was in generating proofs that can be checked
efficiently.  \Drattrim{}, the standard
checker for DRAT proofs~\cite{wetzler14_drattrim}, cannot handle the large proofs
generated by \tbuddy{}, and especially their heavy use of extension
variables.  Fortunately, we were able to achieve acceptable
performance with a checker for the hybrid FRAT proof
format~\cite{baek:tacas:2021}, where many of the proof steps,
including all steps generated by \tbuddy{}, are annotated with hints,
greatly simplifying the operation of the checker.  Although FRAT
proofs use a different file format than the standard DRAT format, the
associated checker \fratrs{} can generate detailed proofs in the same
LRAT format~\cite{lrat} as does \drattrim{}.  In this way, both checkers can make
use of the variety of available LRAT proof checkers, including several that have
been formally verified~\cite{lrat,Tan:2021}.

This paper describes the
integration of \tbuddy{} into \cms{} and provides an experimental
evaluation using two families of benchmark problems.

\section{Solver/Prover Integration}

For parity constraint $P$ having the form shown in Equation~\ref{eqn:parity},
let $\var(P) = \{x_{i_1}, x_{i_2}, \ldots, x_{i_k}\}$ consist of the
variables in the constraint and $\phase(P) = p$ be the phase.  The
sum of constraints $P_a$ and $P_b$ is also a parity constraint $P_c$,
written as $P_c = P_a \oplus P_b$,
with $\var(P_c) = \var(P_a) \triangle \var(P_b)$ (their symmetric
difference) and $\phase(P_c) = \phase(P_a) \oplus \phase(P_b)$.

Let $\phi$ denote the set of input clauses and $P_1, P_2, \ldots, P_m$
denote the set of parity constraints embedded in the input formula.
For $S \subseteq \{1, 2, \ldots, m\}$, let $\psum(S)$ denote the
parity constraint formed by summing all constraints
$P_i$ such that $i \in S$.  A {\em literal} $\lit$ is either a variable or
its complement, with its {\em phase} being 1 in the former case and 0
in the latter.

When integrated into a CDCL framework, Gauss-Jordan elimination
provides a way to systematically sum parity constraints and thereby
detect conflicts and unit propagations implied by the
constraints~\cite{han:cav:2012,laitinen:TAI:2012}.  Suppose that the
state at some point in the search is characterized by a {\em trace}
given by as a conjunction of literals
$\lit_1 \land \lit_2 \land \cdots \land  \lit_k$.  {\em A conflict} arises when there is a sum $\psum(S)$ such
that 1) the variables in $\psum(S)$ are a subset of the variables occurring
in the trace, and 2) the phase of $\psum(S)$ does not match what would be
obtained by summing the phases of the corresponding trace literals.  A {\em  unit
propagation} arises if all but one of the variables in $\psum(S)$ occurs in
the trace, and the implied literal is either this variable or its
complement, such that the sum of the phases of the literals matches that of the constraint.
Both of these inferences can be characterized by a {\em reason} clause
$C$.  For a conflict, $C$ is the {\em conflict clause}, consisting of
the complements of some subset of the literals in the trace.  In other
words, satisfying $\phi$ requires contradicting at least one
assignment in the trace.  For unit propagation, $C$ consists of the
complements of some of the trace literals plus the literal being
inferred by unit propagation.  In other words, any satisfying
assignment to $\phi$ compatible with the trace must also satisfy the
implied literal.

Adding proof generation for parity constraints to \cms{} requires
three new capabilities: \textbf{1)} generating proof steps justifying that the initial parity
  constraints are encoded in the input formula, \textbf{2)} tracking the subset of parity constraints $S$ that give rise to each reason clause $C$, and
\textbf{3)} generating proof steps to justify reason clause $C$ based on those constraints.

The first and third capabilities are provided by \tbuddy{}, while
the second involves adding a tracking capability to \cms{}.

\subsection{Trusted BDDs (TBDDs)}

The \tbuddy{} package supports {\em trusted} BDDs (TBDDs) as its core
data structure.  A trusted BDD represents a Boolean function that is
guaranteed to yield 1 for any assignment that satisfies the input
formula.  Each trusted BDD $\trust{\nodeu}$ is represented by 1) root
node $\nodeu$ in the BDD data structure, 2) extension variable $u$
associated with BDD node $\nodeu$, and 3) a series of proof steps
leading to a step for the unit clause $[u]$, indicating that any
assignment that satisfies the formula must also assign \tautology{} to
$u$.  Performing a sequence of TBDD operations leading to the
generation of trusted BDD $\tleafzero$, where $\leafzero$ is the
terminal BDD node for constant $\nil$, has the effect of generating a
sequence of proof steps leading to the empty clause.
\Tbuddy{} supports several key operations on TBDDs~\cite{bryant:fmcad:2022}:
\begin{description}
\item [{conjunction}] has argument TBDDs
$\trust{\nodeu}$ and $\trust{\nodev}$.  It generates a TBDD
$\trust{\nodew}$ representing the conjunction of the functions
represented by BDDs $\nodeu$ and $\nodev$.

\item [upgrade]  has as arguments trusted BDD
  $\trust{\nodeu}$ and regular BDD $\nodev$.  It upgrades $\nodev$ to
  TBDD $\trust{\nodev}$ based on a proof that $u \rightarrow v$.

\item [clause justification]  has as arguments trusted BDD
  $\trust{\nodeu}$ and clause $C$.  It generates one or more proof
  steps leading to a step consisting of clause $C$ based on the
  implication $u \rightarrow C$.
\end{description}

\subsection{Justifying the Initial Parity Constraints}

\Cms{} uses heuristic methods to detect parity constraints encoded in
clausal form.  When it detects an input parity constraint $P_i$ over
$k$ variables, it has \tbuddy{} generate a TBDD representation
$\trust{\nodeu}_i$ of $P_i$.  To do so, \tbuddy{} simply adds the
$2^{k-1}$ clauses encoding $P_i$ to the proof.  Each of these should
follow from some subset of the input clauses by reverse unit
propagation (RUP), and therefore the proof checker can add the
necessary hints to the proof.  Based on these, \tbuddy{} uses its
conjunction operation to form a TBDD $\trust{\nodev}$ representing the
conjunction of the added clauses.  It also directly generates a BDD
representation $\nodeu_i$ of $P_i$, having the form shown in
Figure~\ref{fig:parity:bdd}, but with the specified set of variables
and the specified phase.  It then uses $\trust{\nodev}$ to upgrade
$\nodeu_i$ to TBDD $\trust{\nodeu}_i$.  In most cases, $\nodeu$ and
$\nodev$ are identical, and so the implication is a tautology.

\subsection{Tracking Constraint Origins}

The standard implementation of G-J elimination in
\cms{}~\cite{DBLP:conf/cav/SoosGM20} need not track which of the original parity
constraints contribute to a conflict or propagation. Instead, it can
directly derive the reason clause from its {\em parity matrix}, the representation of parity constraints it maintains during Gauss-Jordan elimination. In order to add tracking, the modified code keeps a binary
{\em shadow matrix} $M$, having dimension $n \times n$ for a parity matrix with $n$ rows.
At startup, $M$ is initialized to the identity
matrix.
The shadow matrix, as its name implies, shadows the operations on
the parity matrix. When two rows are swapped in the parity matrix,
the shadow matrix also swaps rows, and when two rows are summed,
the shadow matrix also sums the corresponding two
rows. Hence, at all times, entry $i,j$ of $M$ is set to 1 under the condition that, in forming what is now constraint $i$,
initial constraint $j$ was added an odd number of times.
(Observe that summing a constraint an even number of times effectively cancels it out.)
With the shadow matrix, the program can determine the origin of a
row of the parity matrix when it causes conflict or unit propagation.
The shadow matrix can be stored and manipulated with the same
bit-vector representation used for the parity matrix.

\textbf{Example.} Let us take as initial parity constraints $\{P_1:=x_1\oplus x_2 = 1, P_2:=x_1\oplus x_3 = 0, P_3:=x_1\oplus x_2\oplus x_3 = 1\}$.  Then the initial matrix $M_0$ and corresponding shadow matrix $S_0$ is as follows. Notice that the last column of $M_0$ contains the phase values. Since $M_0$ has 3 rows, $S_0$ is the $3 \times 3$ identity matrix:
$$
M_0 =
  \left[
  \begin{array}{ccc|c}
  	x_1 & x_2 & x_3 &\\
    1 & 1 & 0 & 1\\
    1 & 0 & 1 & 0\\
    1 & 1 & 1 & 1\\
  \end{array}
  \right]
\qquad
S_0 =
  \begin{bmatrix}
    P_1 & P_2 & P_3\\
    1 & 0 & 0\\
    0 & 1 & 0\\
    0 & 0 & 1\\
  \end{bmatrix}
$$
The G-J elimination algorithm now takes the 1st row of $M_0$, and sums it into the 2nd and 3rd rows of $M_0$ in order to eliminate the non-zero values in the first column of $M_0$ for all but the first row. Let's call the resulting matrices $M_1$ and $S_1$:
$$
M_1 =
  \left[
  \begin{array}{ccc|c}
    x_1 & x_2 & x_3\\
    1 & 1 & 0 & 1\\
    0 & 1 & 1 & 1\\
    0 & 0 & 1 & 0\\
  \end{array}
  \right]
\qquad
S_1 =
  \begin{bmatrix}
    P_1 & P_2 & P_3\\
    1 & 0 & 0 \\
    1 & 1 & 0 \\
    1 & 0 & 1 \\
  \end{bmatrix}
  $$
  Notice that $S_1$'s 2nd row now reads $1 1 0$, i.e. $M_1$'s 1st row
  can be reconstructed via $P_1\oplus P_2$. Similarly, $S_1$'s 3rd row
  reads $1 0 1$, i.e. $M_1$'s 3rd row can be reconstructed via
  $P_1\oplus P_3$. In this way, by mirroring row swap and row sum
  operations, $S_n$ can always be used to read out which original
  parity constraints need to be summed obtain the corresponding
  constraint in $M_n$.




\subsection{Justifying Reason Clauses}

Each reason clause $C$ must be justified based on a subset $S$ of the
initial parity constraints.  This involves forming a TBDD
representation $\trust{\nodev}$ of the sum constraint $\psum(S)$ and
then using $\trust{\nodev}$ to justify $C$.  Given TBDDs
$\trust{\nodeu}_a$ and $\trust{\nodeu}_b$ representing parity
constraints $P_a$ and $P_b$, \tbuddy{} can generate a TBDD
representation $\trust{\nodeu}_c$ of parity constraint $P_c = P_a
\oplus P_b$.  It does so by the following steps.  First, it conjuncts
$\trust{\nodeu}_a$ and $\trust{\nodeu}_b$ to form TBDD
$\trust{\nodew}$.  Then, it generates the BDD representation $\nodeu_c$ for $P_c$, having
the same form as shown in Figure~\ref{fig:parity:bdd}.  Finally, it
upgrades $\nodeu_c$ to TBDD $\trust{\nodeu}_c$ via a proof of the
implication $w \rightarrow u_c$.

The BDD representation $\nodew$ of
the conjunction of constraints $P_a$ and $P_b$ can be of size
$\Theta(k_a \cdot k_b)$, where $k_a$ and $k_b$ are the number of
nonzero coefficients in $P_a$ and $P_b$, respectively.  On the other
hand, the BDD representation $\nodeu_c$ of their sum will be of size
$O(k_a + k_b)$.  Since the sizes keep growing over multiple elimination
steps, the resulting savings in BDD operations (and therefore proof steps), can be considerable.

The TBDD representation for the sum of constraints $S$ is generated by
performing pairwise sums.  Since the BDD representation of a
constraint with $k$ variables has $2k-1$ nodes, it is important to
sum the constraints in a way that
preserves sparseness (i.e., keeping the number of variables in the constraints low) among the intermediate results.
\Tbuddy{} does so with a greedy selection policy that at each step
sums the pair of constraints $P_a$ and $P_b$ that minimizes the size
of $\var(P_a) \triangle \var(P_b)$, with ties broken randomly.  This can be done efficiently
using a priority queue and incremental updating.

Finally, the clause justification operation adds the reason clause $C$
to the proof based on TBDD $\trust{\nodev}$, representing the sum of the constraints.
A single RUP step suffices for a reason clause arising from a parity constraint.

Once the TBDD representation of a sum of constraints has been used to justify a reason clause,
it can be dereferenced, allowing the \tbuddy{} garbage
collector to reclaim nodes and to delete some of the proof clauses.
This dereferencing is done only when the row of the parity matrix is
modified.  As a result, a given row can be used to justify multiple
reason clauses.

\section{Experimental Results}

We have performed experimental evaluations for several families of
unsatisfiable formulas that are especially challenging for CDCL
solvers.  We compared different operating modes of \cms{} to
\tbsat{}, a BDD-based solver built on top of \tbuddy{}.  \Tbsat{} has
its own Gauss-Jordan elimination solver, and it can combine parity reasoning with
{\em bucket elimination}~\cite{dechter-ai-1999,Jussila:2006,pan-sat-2004}, a
systematic way to reduce a set of BDDs to one of the two possible
terminal nodes via a sequence of conjunctions and
quantifications~\cite{bryant:arxiv:2021,bryant:fmcad:2022}.

We found that the standard \drattrim{} proof checker performs poorly
on proofs generated by \tbuddy{}, due to the large number of RAT
lemmas used to define extension variables.  Instead, we modified
\cms{} to generate proofs in FRAT format~\cite{baek:tacas:2021}, with
over 50\% of the proof clauses generated by \cms{} and 100\% of those
generated by \tbuddy{} having hints.  The associated
\fratrs{} verifier can make use of these hints, while also generating hints for
the other clauses.

\subsection{Urquhart Formulas}

Urquhart's formulas~\cite{Urquhart:1987} consist of sets of parity
constraints based on undirected graphs in a manner similar to a
construction of Tseitin~\cite{Tseitin:1983}.  There is a variable
associated with each edge, and the formula contains an even or odd
parity constraint for each node over its incident edges.  Each edge
occurs in two constraints, and so the formula will be unsatisfiable as
long as the sum of the node phases is odd.  We used a benchmark
generator  written by Li~\cite{li-dam-2003}
that is parameterized by both the graph size $m$ and a value $p$
with $25 \leq p \leq 75$ indicating the percentage of nodes that are
assigned odd parity.

As mentioned earlier, even a minimum instance of these formulas,
having $m=3$, is beyond the reach of today's CDCL solvers.  Applying
Gaussian elimination to these constraints, on the other hand, yields
the infeasible constraint $0=1$, and so they are readily handled just
using parity constraint reasoning.  We have also shown that a
BDD-based solver has polynomial scaling on these formulas when using
bucket elimination~\cite{bryant:arxiv:2021,bryant:fmcad:2022}.

We generated Urquhart Formulas for hardness parameter $m$ ranging from
$3$ to $15$ and with phase percentage parameter $p$ ranging from $26$
to $70$.  Tests were performed on AMD Ryzen~9~5950X processors with a
time limit of 3600 seconds per run.  The solving and verification
times for the two solvers are shown in Table~\ref{tab:urqu-cms}.  The
rightmost column shows the time required by \fratrs{} to check the
proofs generated by \cms{} plus \tbuddy{}.  As can be seen, the
Urquhart formulas pose no challenge for either solver, and the time to
check the proofs are comparable to their generation times.  \Tbsat{}'s
performance advantage over \cms{} on these benchmarks can be attributed to its use of a
sparse representation for parity constraints and that the problem can
be solved purely by parity reasoning.

\begin{table}[tb]
	\centering
    \newcommand{\ch}[1]{\multicolumn{1}{c}{#1}}
	\begin{tabular}{rrrrrrr}
		\toprule
{$m$} & \multicolumn{2}{c}{\Tbsat{}} & \multicolumn{2}{c}{\lratcheck{}}& \multicolumn{1}{c}{\centering {\cmsshort{}}} & \multicolumn{1}{c}{\centering \fratrs{}} \\
      & \multicolumn{2}{c}{\lrat{} enabled} &  \multicolumn{2}{c}{of \tbsat{}} & \multicolumn{1}{c}{\centering {Gauss-J}} & \multicolumn{1}{c}{\centering {Verify}} \\
      & {Gauss-J} & {Bucket-Elim}  & Gauss-J & Bucket-Elim  & \multicolumn{1}{c}{\centering {with FRAT}} & \multicolumn{1}{c}{\centering {\cmsshort{} Proof}} \\
		\midrule
3  & 0.00 & 0.00 & 0.69  & 0.69 &  0.03&  0.06\\
4  & 0.01 & 0.01 & 0.69  & 0.69 &  0.10&  0.17\\
5  & 0.03 & 0.03 & 0.69  & 0.69 &  0.24&  0.40\\
6  & 0.05 & 0.05 & 0.69  & 0.69 &  0.46&  0.84\\
7  & 0.07 & 0.07 & 0.69  & 0.69 &  0.87&  1.66\\
8  & 0.10 & 0.10 & 0.69  & 0.69 & 1.44&  2.96\\
9  & 0.13 & 0.13 & 0.69  & 0.69 & 2.08&  4.34\\
10 & 0.16 & 0.17 & 0.69  & 0.69 &  3.33&  7.30\\
11 & 0.20 & 0.21 & 0.69  & 0.69 &  5.37& 12.26\\
12 & 0.25 & 0.26 & 0.69  & 0.69 &  5.00& 11.73\\
13 & 0.31 & 0.30 & 0.69  & 0.69 &  7.25& 16.61\\
14 & 0.37 & 0.36 & 0.69  & 0.69 & 10.36& 23.96\\
15 & 0.43 & 0.43 & 0.69  & 0.69 & 14.10& 34.47\\

\bottomrule
	\end{tabular}
	\caption{Average times to solve and to verify Urquhart Formulas with \Tbsat{} and \cms{}+\tbuddy{}}
	\label{tab:urqu-cms}
\end{table}

\subsection{Learning Parity with Noise}

Learning Parity with Noise (LPN), also known as the minimal
disagreement parity problem, is a well known hard problem for CDCL SAT
solvers. Crawford contributed several instances of formulas for this
problem as part of the original SATLIB benchmark
suite~\cite{crawford:lpn}.  We wrote a new generator for this problem,\footnote{Available
at \hcref{https://github.com/rebryant/mdp-benchmark}.}
since Crawford's generator is no longer available.  From a
satisfiability solving perspective, these formulas combine a set of
parity constraint with a cardinality constraint. As a consequence,
Gauss-Jordan elimination does not lead directly to an infeasible
constraint, but it can serve a useful role in reasoning about the
parity constraints, while the clausal reasoning component deals with
the cardinality constraint.

The problem is parameterized by values $n$, $m$, and $k$.  There are
$m$ parity constraints defined over a set of $n$ {\em solution
  variables} $s_1, s_2, \ldots, s_n$, plus a set of $m$ {\em
  corruption variables} $r_1, r_2, \ldots, r_m$.  Each parity
constraint $P_i$ depends on some randomly chosen subset of the
solution variables, plus corruption variable $r_i$.  An instance is
generated so that for a {\em target solution}, consisting of a
randomly generated bit sequence $a_1, a_2, \ldots, a_n$,
a subset of
$n-k$ of the constraints  will be satisfied when the solution variables
are set to the target solution and with their associated corruption variables $r_i$ set
to 0, while $k$ of them require $r_i=1$ for the target
solution to satisfy the constraint.  That is, $k$ of the equations
have been ``corrupted'' by flipping their phases.  An at-most-$k$
constraint is placed on the corruption variables, and thus the target
solution will satisfy the formula, as can other solutions, as long as
at most $k$ constraints are corrupted.  The standard LPN formulas are
satisfiable, but by imposing the condition that at most $k-1$ of the
corruption bits can be set to 1, the formula will generally (but not
necessarily) become unsatisfiable.

For each value of $n$ from $20$ to $35$, we generated 20 instances of
LPN formulas, setting $m=2n$.  Each constraint in each formula was
corrupted at random with probability $0.125$, causing the number of
corrupted constraints $k$ in each of the formulas to vary, but having
average value $n/4$.  These choices match the parameters suggested by
Crawford. The formulas were then passed to both \cms{} and \tbsat{} to
be solved, and the resulting proofs were respectively passed to
\lratcheck{} and \fratrs{} to verify.

For \Cms{}, we ran these on
Intel~2xE5-2690v3 CPUs with 24GB RAM allocated for each run, a time
limit of 3600 seconds. For \tbsat{}, we ran these on a Ryzen 9 5950x with 4GB allocated for each run, a time limit of 3600s, and a proof size limit of $2^{30}$ clauses.  The proof size limit was set based on our experience with the \lratcheck{} checker.  Allowing larger proofs would only generate unwieldy files and create proofs that are too large to be checked.

\begin{table}[tb]
\small
	\centering
\begin{tabular}{rrrrrrrr}
\toprule
\makebox[5mm]{$n$} & \makebox[10mm]{UNSAT} &
\multicolumn{2}{c}{\cms{}} &
\makebox[26mm]{\cms{}} &
\multicolumn{2}{c}{\cms{}} &
\makebox[20mm]{\fratrs{}}      \\
 &  & \multicolumn{2}{c}{without G-J} & \makebox[26mm]{with G-J}& \multicolumn{2}{c}{with G-J+\tbuddy} & \makebox[20mm]{avg (s)}\\

&  &
\makebox[10mm]{T.O.} & \makebox[16mm]{PAR-2} &
\makebox[26mm]{no verification} &
\makebox[10mm]{T.O.} & \makebox[16mm]{PAR-2}
& \\
\midrule
20&   15  &  0 &    4.59 &  0.03 &  0 &   1.00     &    0.94\\
21&   10  &  0 &    4.29 &  0.03 &  0 &   1.06     &    1.64 \\
22&   13  &  0 &   20.67 &  0.05 &  0 &   2.41     &    3.98\\
23&   14  &  0 &   31.96 &  0.05 &  0 &  2.79     &    2.55\\
24&   16  &  0 &  113.09 &  0.05 &  0 &   3.40     &    4.51\\
25&   11  &  0 &  257.20 &  0.14 &  0 &   9.45     &   20.51\\
26&   14  &  0 &  908.71 &  0.25 &  0 &  17.70     &   28.63\\
27&   13  &  5 & 2401.46 &  0.31 &  0 &  22.91     &   42.79\\
28&   8   &  5 & 2184.10 &  0.73 &  0 &  38.39     &   42.95 \\
29&   8   & 10 & 3800.17 &  1.27 &  0 &  70.64     &   159.45 \\
30&   14  & 12 & 4696.38 &  1.70 &  0 &  101.98    &   289.13 \\
31&   15  & 13 & 5120.66 &  3.07 &  0 &  194.07    &   726.44 \\
32&   12  & 15 & 5519.15 &  3.09 &  0 &  216.26    &   553.69 \\
33&  10   & 16 & 6167.31 &  6.71 &  0 &  364.75    &   1996.44  \\
34&   12  & 18 & 6631.76 & 19.99 &  5 &  2169.01   &   1391.08  \\
35&   9   & 18 & 6493.04 & 14.79 &  5 &  2144.69   &   1922.08  \\
		\bottomrule
	\end{tabular}
	\caption{Applying different configurations of \cms{} to LPN formulas.  There were 20 instances for each value of $n$.
 The verification times are shown only for UNSAT instances.
 Columns labeled ``T.O.'' indicate the number of cases that exceeded the 3600-second time limit.}
	\label{tab:lpn-cms-par2}
\end{table}

\begin{table}[tb]
\small
	\centering
\begin{tabular}{rrrrr}
\toprule
\multicolumn{1}{c}{\centering {N}}  &
\multicolumn{1}{c}{\centering {UNSAT}}  &
\multicolumn{2}{c}{\centering {\tbsat{} G-J}} &
\multicolumn{1}{c}{\centering {\lratcheck{}}}\\


  &
  &
\multicolumn{1}{c}{\centering {P.O.}} &
\multicolumn{1}{c}{\centering {PAR-2}} &
\multicolumn{1}{c}{\centering {avg (s)}}\\

\midrule
20 &   15 &  0 & 0.18     & 0.38  \\
21 &   10 &  0 & 0.55     & 0.22  \\
22 &   13 &  0 & 0.37     & 0.14  \\
23 &   14 &  0 & 0.43     & 0.11  \\
24 &   16 &  0 & 0.36     & 0.13  \\
25 &   11 &  0 & 4.49     & 0.17  \\
26 &   14 &  0 & 1.54     & 0.21  \\
27 &   13 &  0 & 16.19    & 0.25   \\
28 &   8  &  0 & 42.51    & 0.25   \\
29 &   8  &  0 & 69.72    & 0.44   \\
30 &   14 &  2 & 756.12   & 0.60   \\
31 &   15 &  1 & 387.51   & 0.69   \\
32 &   12 &  3 & 1104.10  & 0.64   \\
33 &   10 &  2 & 792.41   & 1.15   \\
34 &   12 &  4 & 1489.05  & 3.24   \\
35 &   9  &  4 & 1565.19  & 2.32   \\
		\bottomrule
	\end{tabular}
\caption{Applying \tbsat{} in Gauss-Jordan mode to LPN formulas.  There were 20 instances for each value of $n$. The verification times are shown only for UNSAT instances. The column labeled ``P.O.'' indicates the number of instances for which the generated proof exceeded $2^{30}$ clauses.}
	\label{tab:lpn-tbsat}
\end{table}

\paragraph{\Cms{} and \fratrs{} performance}
Table~\ref{tab:lpn-cms-par2} shows the performance of several
different ways to apply \cms{} to these formulas, with the associated PAR-2 scores\footnote{PAR-2 scores are used in the SAT competitions to measure performance. Each benchmark contributes a score that is the number of seconds used to solve it, or in case of a timeout or memory out, twice the timeout in seconds. The average score for all benchmarks is then calculated, giving PAR-2.}.  As is shown, the
number of unsatisfiable formulas for each value of $n$ ranges from
8 (40\%) to 16 (80\%).  Without Gauss-Jordan elimination, \cms{} starts hitting
the time limit with $n=27$ and does so for 90\% of the formulas once
$n$ reaches 34. As a result, the PAR-2 times trend toward the limiting
value of 7200.  On the other hand, using Gauss-Jordan elimination and
without concern for proof generation, all of the benchmarks could
be solved in under 20 seconds.

The measurements for \cms{} in combination with \tbuddy{} were
performed by running the solver with proof generation enabled,
regardless of whether or not the formula was satisfiable.
The combination
enables proof generation for much larger problems than is possible with pure CDCL\@.
The solver exceeded the time limit
only for $n\geq 34$.  
Compared to
having no proof generation, it exacts a significant performance
penalty.  The increased time
could be partly due to the loose integration of our implementation.
\Tbuddy{} must start fresh in summing a subset of the parity
constraints every time a new parity matrix row
is used to generate a reason clause,
even if many elements of the subset have been summed previously.
A tighter integration, however, would require major rewriting of the already-complex parity reasoning code.

The final column shows the average times for running \fratrs{} for the
cases where the formula is unsatisfiable and \cms{} is able to
complete its execution.  Overall, the performance is acceptable, but
\fratrs{} can take around $3\times$ more time to check a proof than
was required to generate it.  In these proofs, \tbuddy{} provides
hints for each of its proof steps, and \cms{} does so for over 50\% of
them, and so it seems like it should be possible to improve the
checker performance.

\paragraph{\Tbsat{} and \lratcheck{} performance}
Table~\ref{tab:lpn-tbsat} shows the performance of \tbsat{} with
Gauss-Jordan elimination and \lrat{} proof generation applied to the
LPN formulas.  As before, the number of unsatisfiable formulas for
each value of $n$ ranges from 8 (40\%) to 16 (80\%).  A separate
program detects parity constraints in the CNF representation and
generates a {\em schedule} for \tbsat{} identifying each constraint
and the set of clauses that give rise to it~\cite{bryant:tacas:2022}.
\Tbsat{} lacks any variable ordering heuristics, and so
the user must supply a variable ordering to achieve
good performance on these formulas.  Our ordering for these formulas
is as follows: the variables encoding the at-most-one constraints, the
corruption variables, the auxilliary variables for the parity
constraints, and finally the solution variables.

Overall, \Tbsat{} ran significantly faster than \cms{} on these
formulas, even accounting for the performance difference of the
execution platforms.  On the other hand, it was less stable.  Even for
formulas as small as $n=30$, it could encounter a single conjunction
operation that would cause so many BDD operations that the proof
generator exceeded the limit of $2^{30}$ clauses.  By contrast, \cms{}
performance was very stable until it started exceeding the time limit
once $n$ reached 34.  The generated proofs, however, could reach
100~GB for both solvers.  Checking the LRAT proofs generated by
\tbsat{} ran much faster than did the checking of the FRAT proofs
generated by \cms{}.  All of the problems that \tbsat{} could solve
without overflowing the proof clause counter ran in under 1200s.
Indeed, the PAR-2 scores would improve if the timeout limit were
reduced.

\section{Conclusion}

CDCL solvers have a major weakness in their handling of parity
constraints.  Augmenting CDCL with Gauss-Jordan elimination is well
established, but this capability has so far lacked the ability to
generate proofs of unsatisfiability.  Our work shows that standard
clausal proof frameworks suffice, with BDDs providing a bridge between
parity reasoning and proof generation.
We hope that others will explore the
CDCL(T) framework, while maintaining the important capability of
generating unsatisfiability proofs.

\bibliography{references}

\begin{thebibliography}{10}

\bibitem{baek:tacas:2021}
S.~Baek, M.~Carneiro, and M.~J.~H. Heule.
\newblock A flexible proof format for {SAT} solver-elaborator communication.
\newblock In {\em Tools and Algorithms for the Construction and Analysis of
  Systems (TACAS), Part I}, volume 12651 of {\em LNCS}, pages 59--75, 2021.

\bibitem{barnett:cade:2021}
L.~A. Barnett and A.~Biere.
\newblock Non-clausal redundancy properties.
\newblock In {\em Conference on Automated Deduction (CADE)}, volume 12699 of
  {\em LNAI}, pages 252--272, 2021.

\bibitem{biere-kissat-2020}
A.~Biere, K.~Fazekas, M.~Fleury, and M.~Heisinger.
\newblock {CaDiCaL}, {Kissat}, {Paracooba}, {Plingeling} and {Treengeling}
  entering the {SAT Competition 2020}.
\newblock In {\em Proc.~of {SAT Competition} 2020 -- Solver and Benchmark
  Descriptions}, volume B-2020-1 of {\em Department of Computer Science Report
  Series B}, pages 51--53. University of Helsinki, 2020.

\bibitem{bryant:fmcad:2022}
R.~E. Bryant.
\newblock {TBUDDY}:, a proof-generating {BDD} package.
\newblock In {\em Formal Methods in Computer-Aided Design}, 2022.

\bibitem{bryant:tacas:2022}
R.~E. Bryant, A.~Biere, and M.~J.~H. Heule.
\newblock Clausal proofs for pseudo-{B}oolean reasoning.
\newblock In {\em Tools and Algorithms for the Construction and Analysis of
  Systems (TACAS)}, LNCS, 2022.

\bibitem{bryant:tacas:2021}
R.~E. Bryant and M.~J.~H. Heule.
\newblock Generating extended resolution proofs with a {BDD}-based {SAT}
  solver.
\newblock In {\em Tools and Algorithms for the Construction and Analysis of
  Systems (TACAS), Part I}, volume 12651 of {\em LNCS}, pages 76--93, 2021.

\bibitem{bryant:arxiv:2021}
R.~E. Bryant and M.~J.~H. Heule.
\newblock Generating extended resolution proofs with a {BDD}-based {SAT}
  solver.
\newblock {\em CoRR}, abs/2105.00885, 2023.

\bibitem{crawford:lpn}
J.~M. Crawford, M.~J. Kearns, and R.~E. Schapire.
\newblock The minimal disagreement parity problem as a hard satisfiability
  problem.
\newblock Mirrored at
  \url{https://www.cs.cornell.edu/selman/docs/crawford-parity.pdf}, 1994.

\bibitem{demoura:tacas:2008}
L.~{de Moura} and N.~Bj{\o}rner.
\newblock {Z3}: An efficient {SMT} solver.
\newblock In {\em Tools and Algorithms for the Construction and Analysis of
  Systems}, volume 4963 of {\em LNCS}, pages 337--340, 2008.

\bibitem{dechter-ai-1999}
R.~Dechter.
\newblock Bucket elimination: A unifying framework for reasoning.
\newblock {\em Artificial Intelligence}, 113(1--2):41--85, 1999.

\bibitem{gocht:veripb:2020}
S.~Gocht, C.~McCreesh, and J.~{Nordstr\"{o}m}.
\newblock {VeriPB}: The easy way to make your combinatorial search algorithm
  trustworthy.
\newblock In {\em From Constraint Programming to Trustworthy AI}, 2020.

\bibitem{gocht:aaai:2021}
S.~Gocht and J.~{Nordstr\"{o}m}.
\newblock Certifying parity reasoning efficiently using pseudo-{B}oolean
  proofs.
\newblock In {\em AAAI Conference on Artificial Intelligence}, pages
  3768--3777, 2021.

\bibitem{han:cav:2012}
C.-S. Han and J.-H.~R. Jiang.
\newblock When {B}oolean satisfiability meets {G}aussian elimination in a
  simplex way.
\newblock In {\em Computer-Aided Verification (CAV)}, volume 7358 of {\em
  LNCS}, pages 410--426, 2012.

\bibitem{lrat}
M.~J.~H. Heule, W.~A. Hunt, M.~Kaufmann, and N.~D. Wetzler.
\newblock Efficient, verified checking of propositional proofs.
\newblock In {\em Interactive Theorem Proving}, volume 10499 of {\em LNCS},
  pages 269--284, 2017.

\bibitem{Jussila:2006}
T.~Jussila, C.~Sinz, and A.~Biere.
\newblock Extended resolution proofs for symbolic {SAT} solving with
  quantification.
\newblock In {\em Theory and Applications of Satisfiability Testing (SAT)},
  volume 4121 of {\em LNCS}, pages 54--60, 2006.

\bibitem{kullman:dam:1999}
O.~Kullmann.
\newblock On a generalization of extended resolution.
\newblock {\em Discrete Applied Mathematics}, 96-97:149--176, 1999.

\bibitem{laitinen:TAI:2012}
T.~Laitinen, T.~Junttila, and I.~Niemel{\"a}.
\newblock Extending clause learning {SAT} solvers with complete parity
  reasoning.
\newblock In {\em International Conference on Tools with Artificial
  Intelligence}, pages 65--72. IEEE, 2012.

\bibitem{li-dam-2003}
C.-M. Li.
\newblock Equivalent literal propagation in the {DLL} procedure.
\newblock {\em Discrete Applied Mathematics}, 130(2):251--276, 2003.

\bibitem{buddy}
J.~Lind-Nielsen.
\newblock {\em {BuDDy}: a Binary Decision Diagram Package}.
\newblock Department of Information Technology, Technical University of
  Denmark, 1996.

\bibitem{pan-sat-2004}
G.~Pan and M.~Y. Vardi.
\newblock Search vs.~symbolic techniques in satisfiability solving.
\newblock In {\em Theory and Applications of Satisfiability Testing (SAT)},
  volume 3542 of {\em LNCS}, pages 235--250, 2005.

\bibitem{philipp:lai:2016}
T.~Philipp and A.~Rebola-Pardo.
\newblock {DRAT} proofs for {XOR} reasoning.
\newblock In {\em Logics in Artificial Intelligence}, volume 10021 of {\em
  LNAI}, pages 415--429, 2016.

\bibitem{ebddres}
C.~Sinz and A.~Biere.
\newblock Extended resolution proofs for conjoining {BDD}s.
\newblock In {\em Computer Science Symposium in Russia (CSR)}, volume 3967 of
  {\em LNCS}, pages 600--611, 2006.

\bibitem{DBLP:conf/cav/SoosGM20}
M.~Soos, S.~Gocht, and K.~S. Meel.
\newblock Tinted, detached, and lazy {CNF-XOR} solving and its applications to
  counting and sampling.
\newblock In {\em Computer Aided Verification (CAV)}, volume 12224 of {\em
  LNCS}, pages 463--484, 2020.

\bibitem{crypto}
M.~Soos, K.~Nohl, and C.~Castelluccia.
\newblock Extending {SAT} solvers to cryptographic problems.
\newblock In {\em Theory and Applications of Satisfiability Testing (SAT)},
  volume 5584 of {\em LNCS}, pages 244--257, 2009.

\bibitem{Tan:2021}
Y.~K. Tan, M.~J.~H. Heule, and M.~O. Myreen.
\newblock cake\_lpr: Verified propagation redundancy checking in {CakeML}.
\newblock In {\em Tools and Algorithms for the Construction and Analysis of
  Systems (TACAS), Part II}, volume 12652 of {\em LNCS}, pages 223--241, 2021.

\bibitem{Tseitin:1983}
G.~S. Tseitin.
\newblock On the complexity of derivation in propositional calculus.
\newblock In {\em Automation of Reasoning: 2: Classical Papers on Computational
  Logic 1967--1970}, pages 466--483. Springer, 1983.

\bibitem{Urquhart:1987}
A.~Urquhart.
\newblock Hard examples for resolution.
\newblock {\em J.ACM}, 34(1):209--219, 1987.

\bibitem{wetzler14_drattrim}
N.~D. Wetzler, M.~J.~H. Heule, and W.~A. Hunt~Jr.
\newblock {DRAT}-trim: Efficient checking and trimming using expressive clausal
  proofs.
\newblock In {\em Theory and Applications of Satisfiability Testing (SAT)},
  volume 8561 of {\em LNCS}, pages 422--429, 2014.

\end{thebibliography}

\end{document}